# Sociotechnical Approach to Enterprise Generative Artificial Intelligence (E-GenAI)

Leoncio Jiménez
TUTELKAN
Talca – Chile
leoncio.jimenez@tutelkan.com

Francisco Venegas
DCC - Universidad de Chile
Santiago –Chile
fvenegas@dcc.uchile.cl

***Abstract -*** *In this theoretical article, a sociotechnical approach is proposed to characterize. First, the business ecosystem, focusing on the relationships among Providers, Enterprise, and Customers through SCM, ERP, and CRM platforms to align: (1) Business Intelligence (BI), Fuzzy Logic (FL), and TRIZ (Theory of Inventive Problem Solving), through the OID model, and (2) Knowledge Management (KM) and Imperfect Knowledge Management (IKM), through the OIDK model. Second, the article explores the E-GenAI business ecosystem, which integrates GenAI-based platforms for SCM, ERP, and CRM with GenAI-based platforms for BI, FL, TRIZ, KM, and IKM, to align Large Language Models (LLMs) through the E-GenAI (OID) model.*
*Finally, to understand the dynamics of LLMs, we utilize finite automata to model the relationships between Followers and Followees. This facilitates the construction of LLMs that can identify specific characteristics of users on a social media platform.*

## I. Introduction

Systems are open when interacting with the environment to maintain their identity, autonomy, and purpose (Rosnay J., 1980). Identity distinguishes the system (enterprise, business process or unit) from other systems. Autonomy refers to the decision-making process of the system to maintain its organization (soul of the enterprise, business process, or business unit) and its structure (body of enterprise, business process, or business unit). The purpose refers to the objectives of the system (enterprise, business process or unit) to maintain the organization (soul) of the structure (body) over time (the structure change). Open systems take in inputs, which are utilized and transformed within the system (such as an enterprise or process), to produce outputs. These outputs can be services or products resulting from the transformation, which are then delivered to the environment. Closed systems are those that lack either inputs or outputs, meaning they do not interact with their environment. All activities occur within the system, and the results of any transformations or operations stay contained inside. A specific type of closed system (organizational closure) is an autopoietic system.

Autopoietic systems are closed systems concerning their organization, but open in their structure. The result of its operation remains in the system. There is no interaction with the environment. People are not autopoietic systems, but rather the nervous system that makes them (Maturana and Varela, Theory of Autopoiesis of Santiago) (Varela, 2007). Enterprises are not autopoietic systems, but the business units (processes) that constitute them (Limone and Bastias, Theory of Autopoiesis of Valparaíso) (Limone & Bastias, 2006).

## II. Sociotechnical Approach

Systems Every organization—whether a family, enterprise, business process, or unit—is, by definition (Bravo, 1998), "an organized group of human beings." Understanding these organizations requires examining their interactions and relationships, as what affects one part can impact the whole both now and in the future. An enterprise (or business process/unit) is defined as a living system that is self-aware and viable through its interactions with its environment (business ecosystem, see Figure 1). To survive and evolve, the enterprise must understand itself, its customers, and its suppliers. Platforms such as supply chain management (SCM), enterprise resource planning (ERP), and customer relationship management (CRM) can assist in this endeavor.

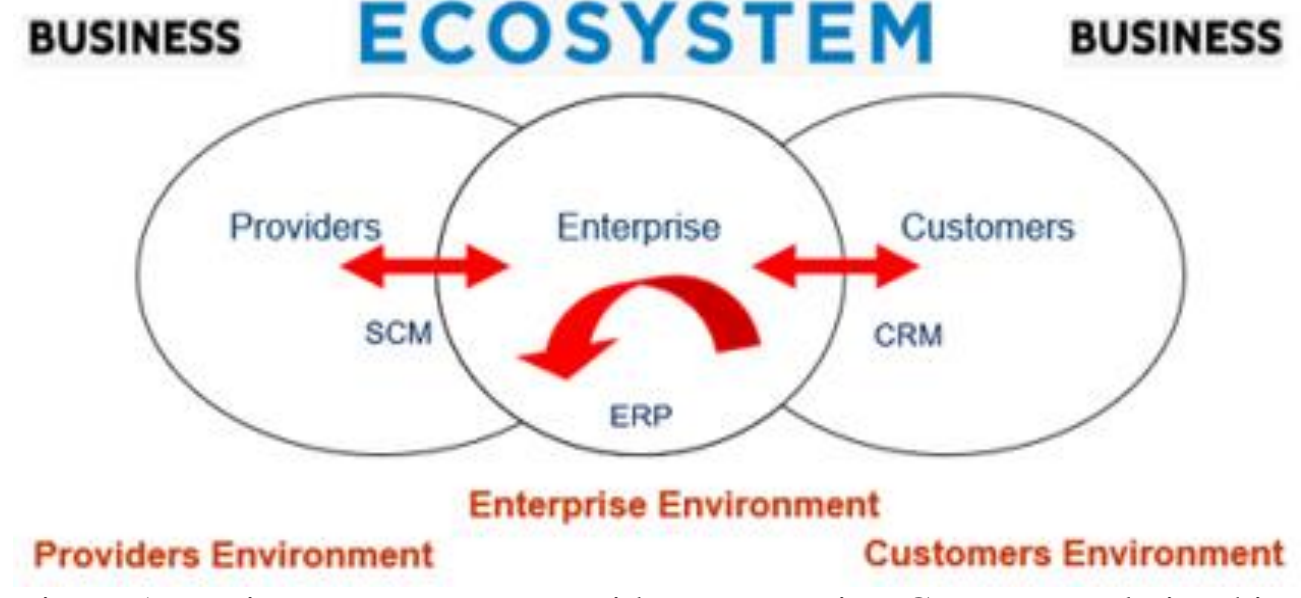

Figure 1: Business Ecosystem. Providers–Enterprise–Customers relationships through SCM, ERP, and CRM platforms. Source: main author.

The enterprise sociotechnical approach necessarily and sufficiently integrates two subsystems that are inseparable from the social and technological context of the enterprise, defined by and for human beings. The social component refers to the needs of people, while the technical component represents the technological imperative to satisfy those needs.

The system is seen as an open system: Inputs–Transformation–Outputs. The system interacts with the environment. Figure 2 shows these relationships and their environment. The social component is defined by Providers–Enterprise–Customers relationships through SCM, ERP, and CRM platforms, that define the technical component of the sociotechnical approach.

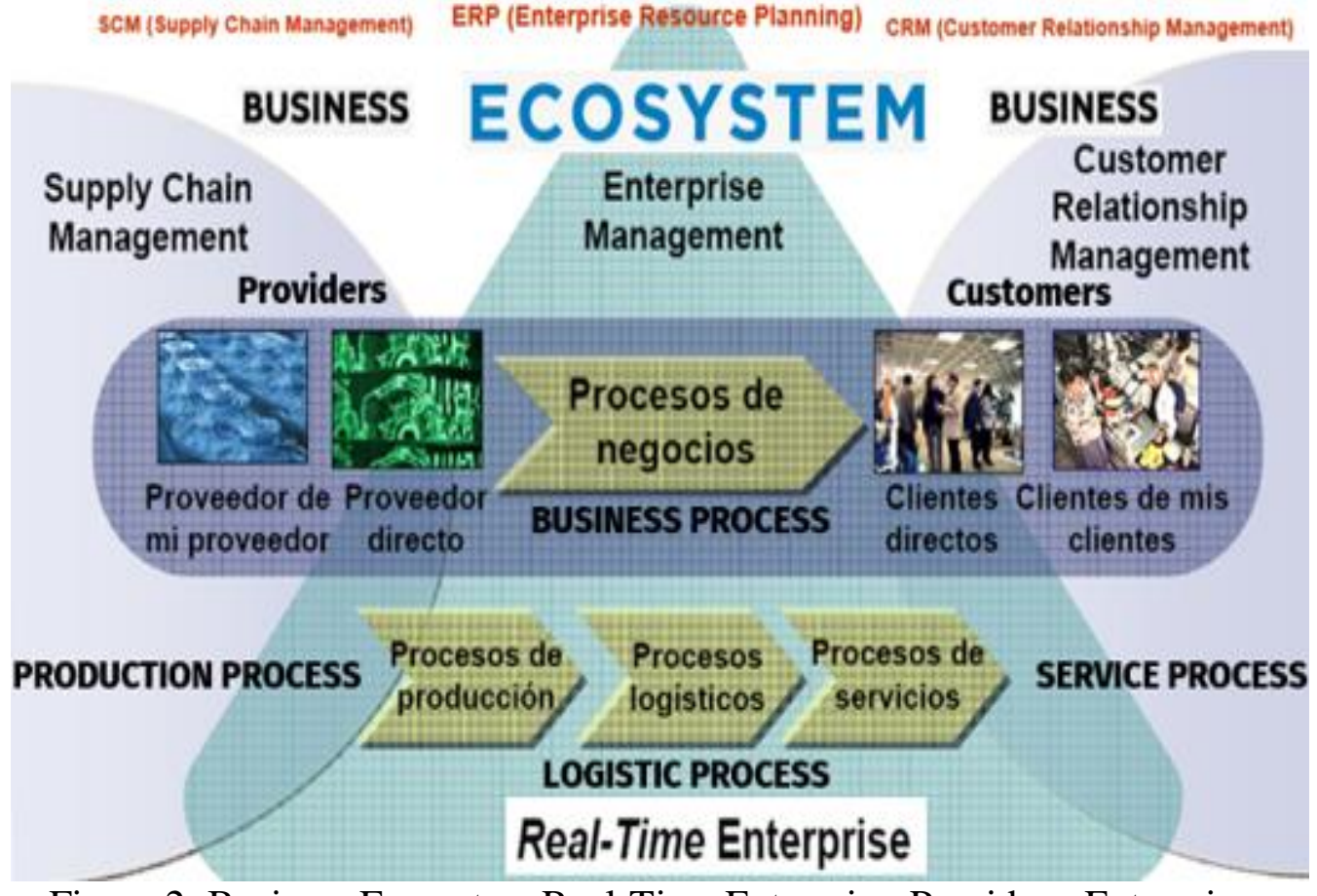

Figure 2: Business Ecosystem Real-Time Enterprise. Providers–Enterprise–Customers relationships through SCM, ERP, and CRM platforms. Source: adapted from (Laengle, 2019).

### A. OID Model

The Le Moigne OID model (see Figure 3) organizes the enterprise (business process or unit) at all time t into three systems: OS operational system, IS information system, and DS decision system (Le Moigne, 1990). These systems interact with each other through variables and values. Figure 3 presents an intervention of Le Moigne OID model, presented in «LA MODELISATION DES SYSTEMES COMPLEXES».

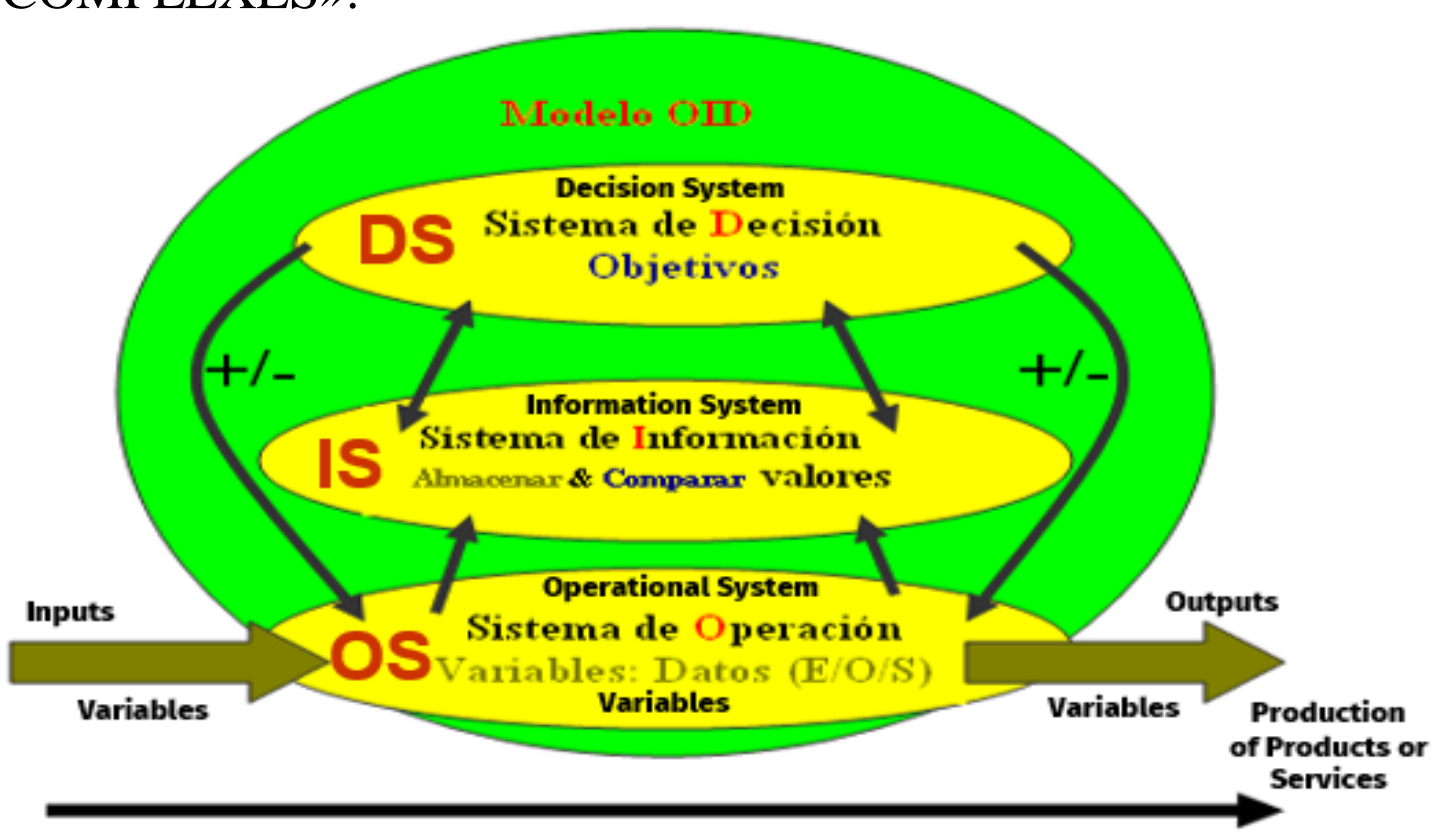

Figure 3: OID model (Operational, Information, Decision). Source: adapted from (Le Moigne, 1990).

At all times t the OS transforms inputs from providers (such raw materials, energy, goods, workflows, and other resources) into outputs to customers (products or services). The IS stores the values of OS's variables and compares them with the target values set by the DS. The DS controls these objectives and adjusts the OS's variable values to ensure alignment with the desired goals.

The OS is an open system, while the IS and DS are considered closed systems. The absence of entry and exit arrows in Figure 3 emphasizes that the IS and DS are closed with respect to their internal organization but open in terms of their structure. The OID model preserves the unity and identity of the system as a whole at all time t. Rather than being viewed as a single system, the OID model should be understood as comprising multiple systems. This is because there are multiple operational systems within the OS, as well as multiple information systems (IS) and decision systems (DS). This is very well explained in «THE EVOLUTION OF ARTIFICIAL INTELLIGENCE TOWARDS AUTONOMOUS SYSTEMS WITH PERSONALITY SIMULATION» by (Colloc, 2019), see Figure 4: "On level 6, the system becomes able to memorize his decision O.I.D (Operating system (OS), Information system (IS), System of Decision (DS)). On level 7, the system coordinates numerous decisions of actions at all the time t, concerning its internal activity, regulation and the external information from and to its environment (SCS). On level 8, the system is endowed with a subsystem of imagination and design (SID). On level 9, the system is able to decide on its decision and to determine the positive and negative aspects of its actions. This finalization of a complex system is close to the human thought (FS) which confers it an autonomy of decision allowing it to set its own goals. Autonomous multiagents systems (AMAS) belongs to this kind Le Moigne (1990)".

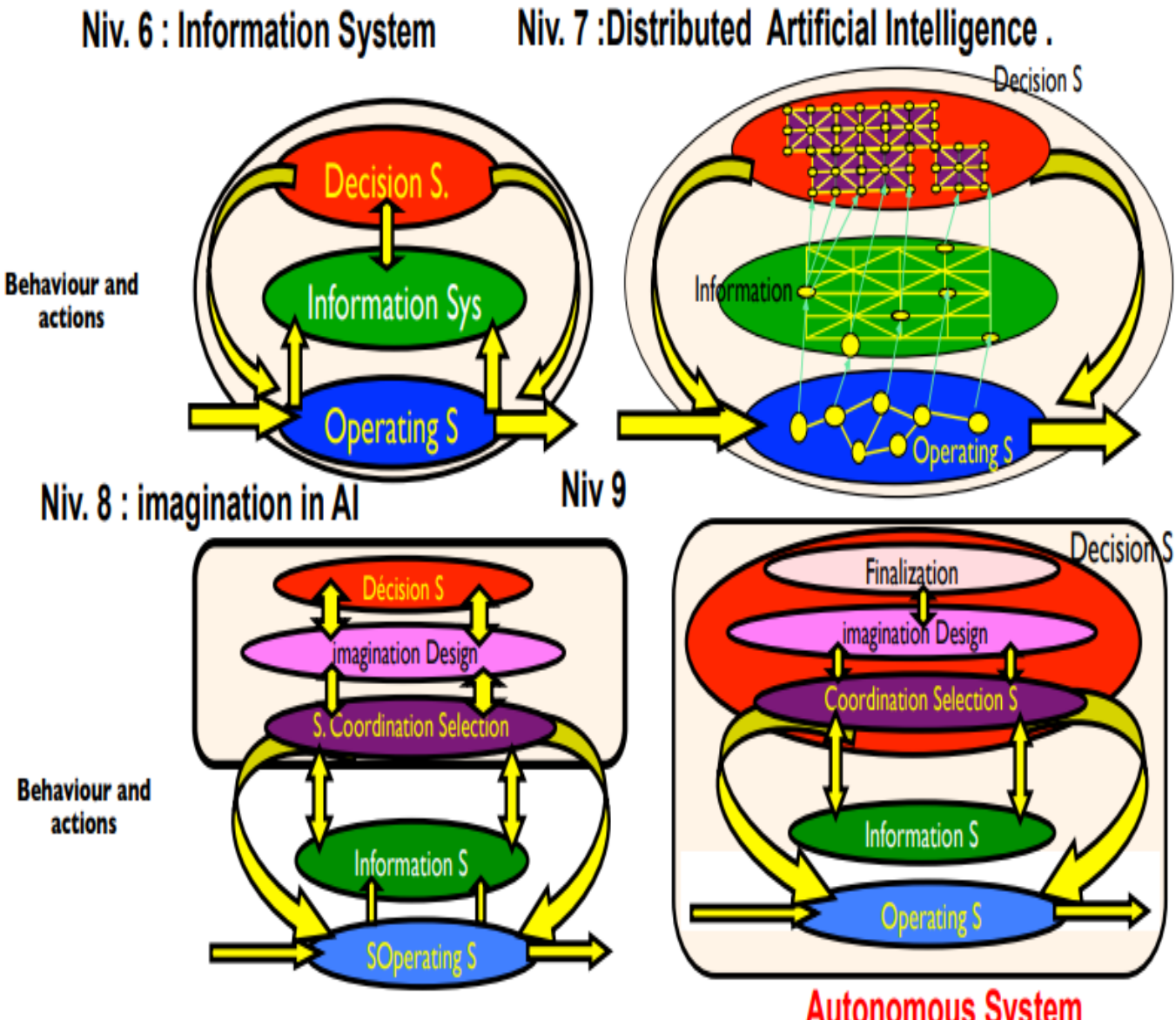

Figure 4: OID model (Operating, Information, Decision). Source: adapted from (Colloc, 2019).

Sigifredo Laengle's perspective on the OID model (see Figure 5) is noteworthy. In this view, the "hand" represents the workflow of the business process or unit, while the "brain" signifies the thinking and decision-making processes involved in converting inputs into outputs. Platforms like CRM, SCM, SRM, and ERP provide real-time visibility into the enterprise (OS) through business models and data systems, which generate information (IS). This information then feeds into dashboard indicators that manage 24/7 the decision-making process (DS). Enterprise 4.0 (the real-time enterprise) is a horizontally integrated system designed to create value within its Providers–Enterprise–Customers relationships (Laengle, 2019).

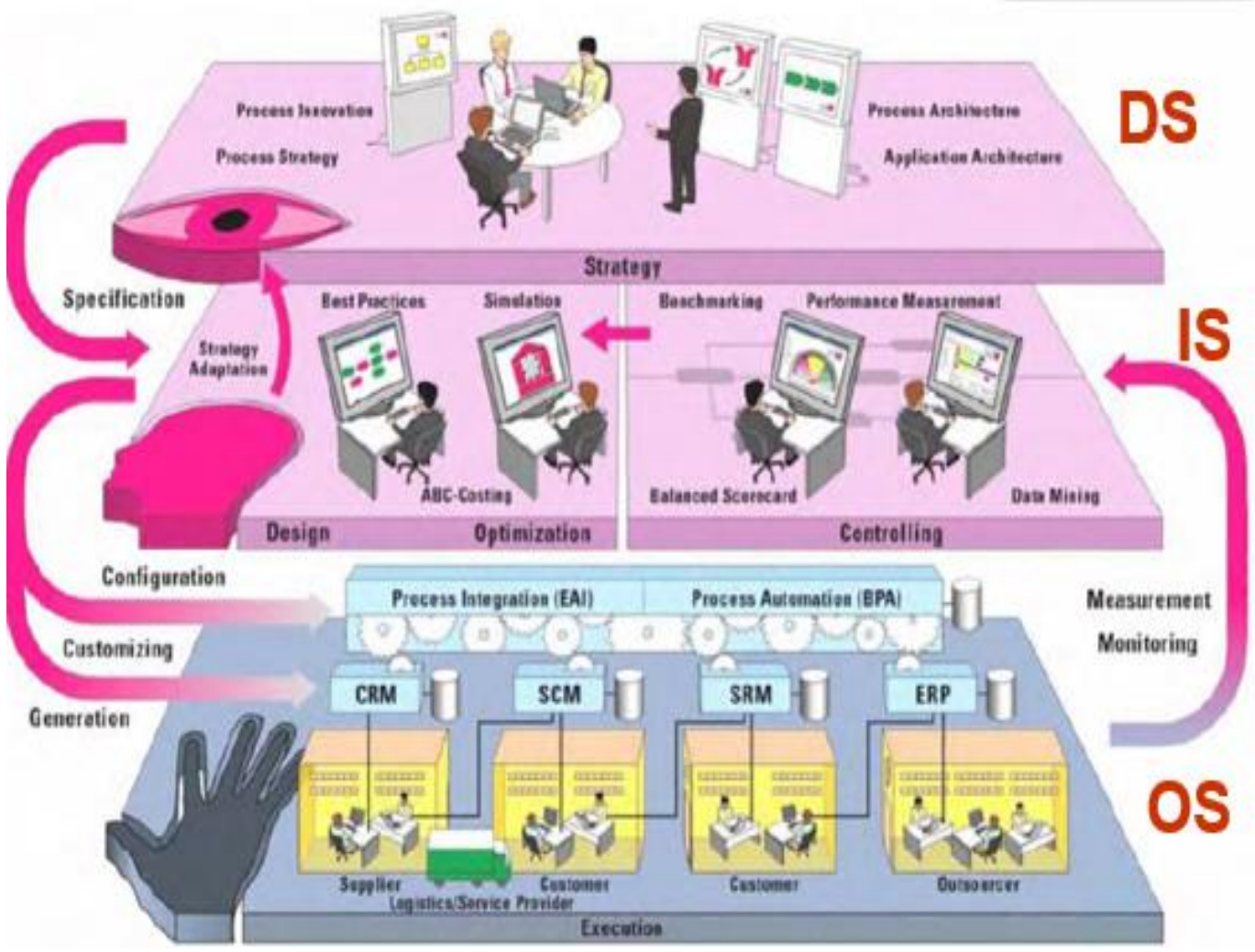

Figure 5: OID model (Operational, Information, Decision). Source: adapted from (Laengle, 2019).

The OS outlines the daily transactions involving operational data flows, including inputs, production (transformation), and outputs. It represents a value chain that encompasses purchasing inputs from suppliers, transforming raw materials, and selling products or services to customers. These data flows (see Figure 6) are regulated through decision and control flows from DS to OS. This regulation involves adjusting the values of OS input variables, operational variables, and output variables. Adjustments are made to align with DS objectives, such as increasing (+) or decreasing (-) raw material purchases for the next month, modifying production levels for the upcoming week, or adjusting product sales for the upcoming holidays. DS manages these variables to meet strategic goals.
The sociotechnical approach that we propose, aligns Business Intelligence (BI), Fuzzy Logic (FL), and TRIZ (Theory of Inventive Problem Solving), through the OID model (see Figure 6).

Business Intelligence (BI) converts IS information flows (see Figure 6) into a management dashboard that displays key indicators. It utilizes both internal information from OS operational systems and external data such as competition, economic trends, government information, and more. The BI dashboard (control panels) monitors essential and critical variable values within the Providers–Enterprise–Customers relationships through SCM, ERP, and CRM platforms (Jiménez & Urrutia, 2013).

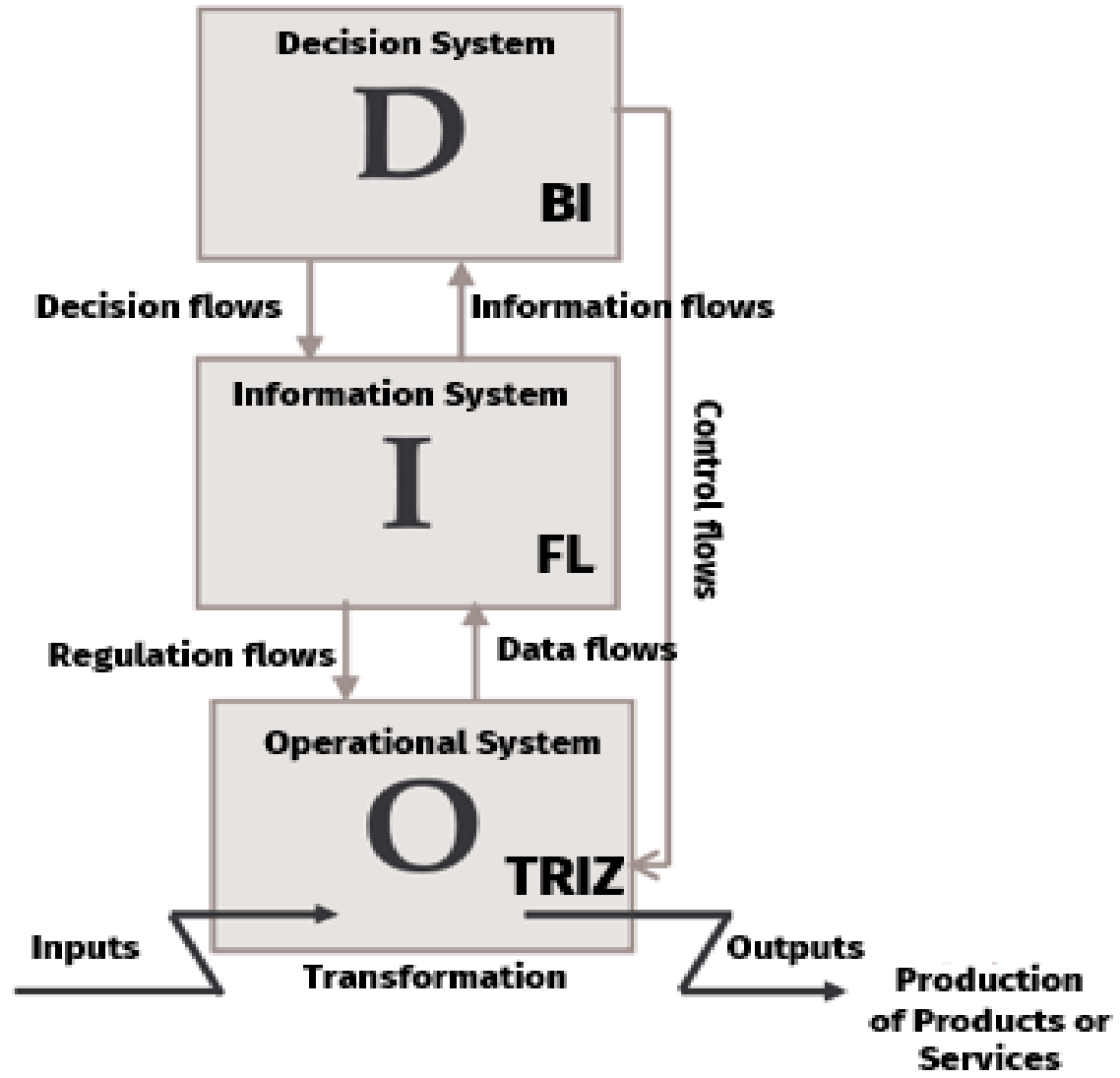

Figure 6: Sociotechnical alignment of BI, FL, and TRIZ in the OID model. Source: adapted from (Ermine, 2014).

Fuzzy Logic (FL) can handle OS data flows (see Figure 6) that involve imprecise, uncertain, or incomplete attributes and values. For instance, it can address scenarios such as water temperatures that are nearly lukewarm, cardboard with slightly wrinkled edges, raw material purchases from an unspecified supplier, uncertain machine operation times without pans, and imprecise quantities of products sold to a customer (Jiménez et al., 2005).

TRIZ (Theory of Inventive Problem Solving) consists of 39 general parameters, 40 inventive principles, 76 standard solutions, 8 evolutionary patterns, and other tools. OS personnel (see Figure 6) can use this methodology to find innovative solutions to their problems. The success or failure of product or service evolution hinges on three factors: (1) the dynamics within the business ecosystem of Providers–Enterprise–Customers, (2) the development of OS, IS, and DS, which together form a complex and interdependent system, where the existence of one relies on the existence of the others. The OID model exists as a unit and loses its identity if one of them fails, and (3) the application of TRIZ in the systematic innovation process to address complex OS issues (Cortes, 2009; 2023); (Jiménez, 2000a; 2000b).

**B. OIDK Model**

The Ermine OIDK model (see Figure 7) introduces a new component, known as the KS (Knowledge System). This component interacts with others through competence flows (which include knowledge related to knowing, knowing doing, knowing being, and others) and cognition flows (which encompass knowledge related to problem-solving, conflict resolution, best practices, creativity, and more) that are accumulated over time by the enterprise's personnel. Competence flows serve as input streams to the KS from OS, IS, and DS, while cognition flows act as output streams from the KS to OS, IS, and DS. The KS functions as a repository for knowledge assets, designed to store and capitalize on the enterprise's daily knowledge (Ermine, 1996; Ermine, 2014).
The sociotechnical approach that we propose, aligns Knowledge Management (KM) and Imperfect Knowledge Management (IKM), through the OIDK model (see Figure 7).

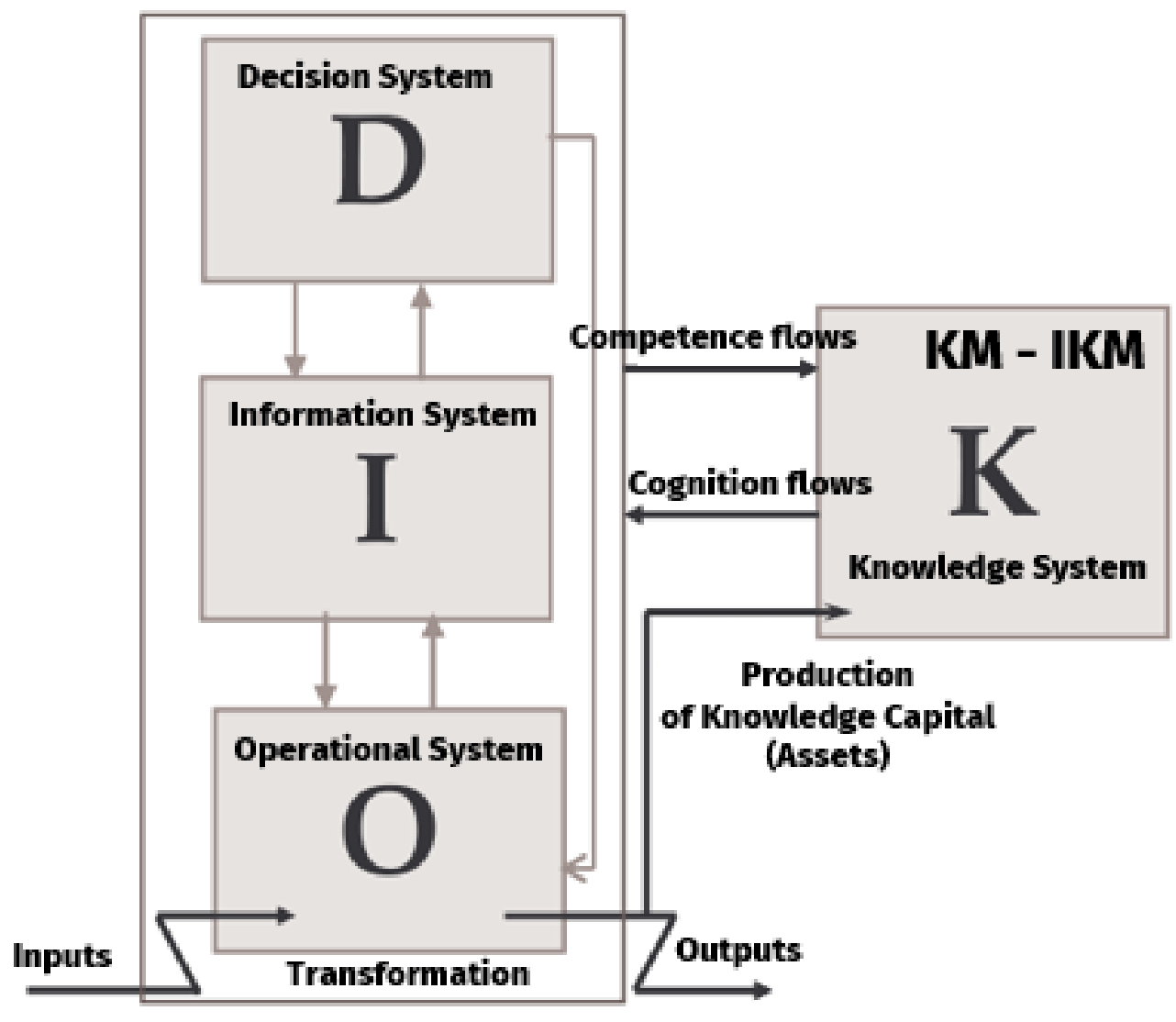

Figure 7: Sociotechnical alignment of KM and IKM in the OIDK model. Source: adapted from (Ermine, 2014).

Knowledge Management (KM) is the process of capitalizing on, storing, and utilizing knowledge assets (see Figure 7) by the personnel in OS, IS, and DS roles during daily business activities or within business units. KM is designed to ensure that knowledge is used effectively while preserving its integrity for future generations of employees (Jiménez, 2007).

Imperfect Knowledge Management (IKM) aids in managing KS competence flows or cognition flows (see Figure 7) by dealing with the imprecise, uncertain, or incomplete aspects of meaning. IKM acknowledges that an enterprise's knowledge is often imperfect, characterized by varying degrees of imprecision, uncertainty, or incompleteness. Additionally, the human aspects of ideas, emotions, and actions that contribute to the creation of new knowledge within an enterprise represent a complex system of human relationships (Jiménez, 2010).

## III. GenAI

Generative Artificial Intelligence (GenAI) represents one of the most exciting and promising branch of Artificial Intelligence (AI) (Feuerriegel et al, 2023). GenAI creates outputs such as text (words and/or code), voice, images, videos, or other data by utilizing Large Language Models (LLMs) and prompt engineering techniques. As illustrated in Figure 8, a GenAI system processes input data and generates corresponding output data. This input and output can be in various forms, including text, voice, images, or videos, and the transformation may involve text-to-text, text-to-image, image-to-text, text-to-audio, or other combinations. The user provides a prompt, such as a question, and LLMs like ChatGPT respond with an answer, which can be correct or incorrect (Banh & Strobel, 2023).

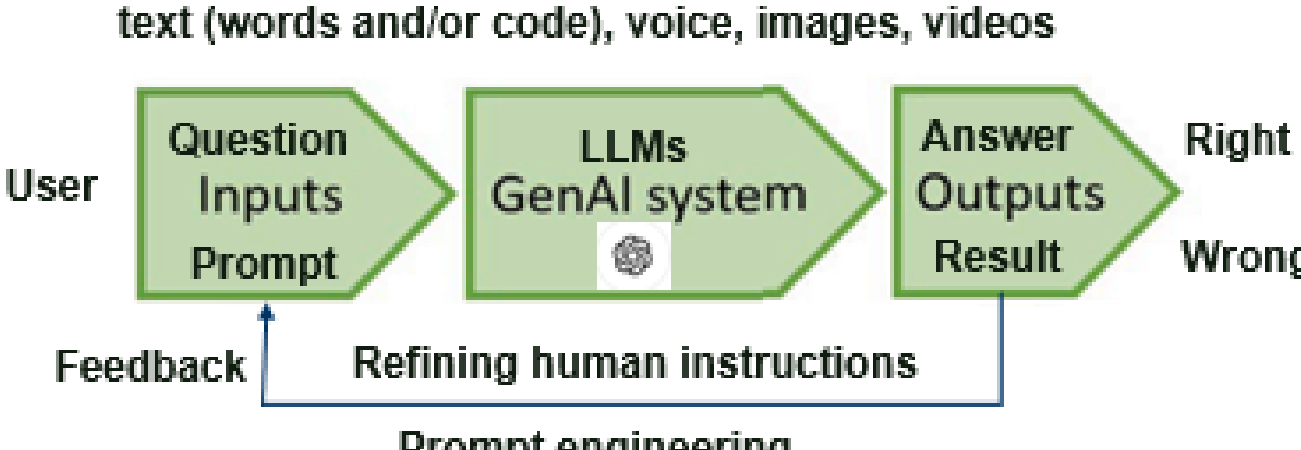

Figure 8: GenAI System. Source: adapted from (Mariani & Dwivedi, 2024).

Prompt engineering involves crafting clear and precise questions or instructions to guide the GenAI system in producing the desired output. If the response is inaccurate, the user must revise the prompt, refining it with additional context or clarity (reshaping the prompt) to achieve the correct result. The feedback loops between machine and human. Effective prompt construction is crucial for GenAI systems to deliver accurate and meaningful answers. Refining human instructions, or prompt engineering, is a critical part of working with Generative AI (GenAI) systems. It involves carefully crafting and adjusting the input data given to the GenAI system to ensure that it produces the most accurate and relevant output data possible. This process requires users to clarify their questions or requests, add context, and break down complex tasks into simpler, more specific prompts. When the GenAI's response is incorrect or incomplete, refining the instructions allows the user to reshape the prompt, providing clearer direction and additional detail. This iterative process is essential for improving the quality of responses and guiding the GenAI towards delivering more accurate and useful results (Mariani & Dwivedi, 2024). GenAI is designed to mimic human creativity and language processing for tasks like content creation, natural language understanding, and conversational interfaces. GenAI "generate" content in the creative or natural language sense. Figure 9 illustrates a prompt GenAI system managed by an orchestration application. When the user provides prompting instructions, the orchestrator may retrieve data from external sources, like databases or the internet, to supply the LLMs with relevant information (Schneider et al., 2024).

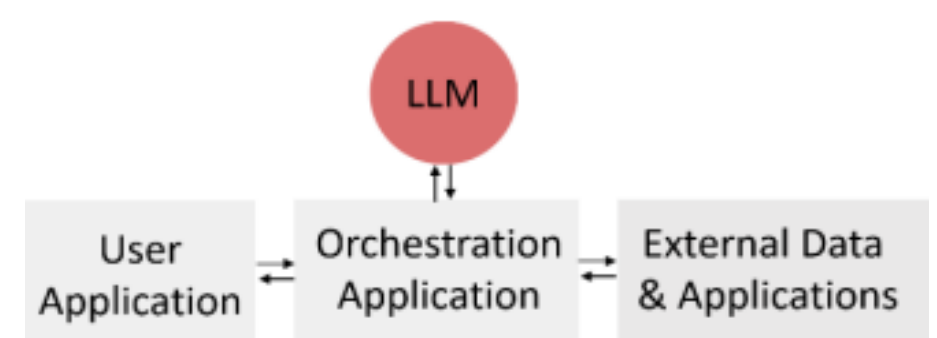

Figure 9: Prompt GenAI System. Source: (Schneider et al., 2024).

Figure 10 shows a prompt GenAI system designed with multiple query instructions to perform the prompt engineering process. (Laskar et al., 2024).

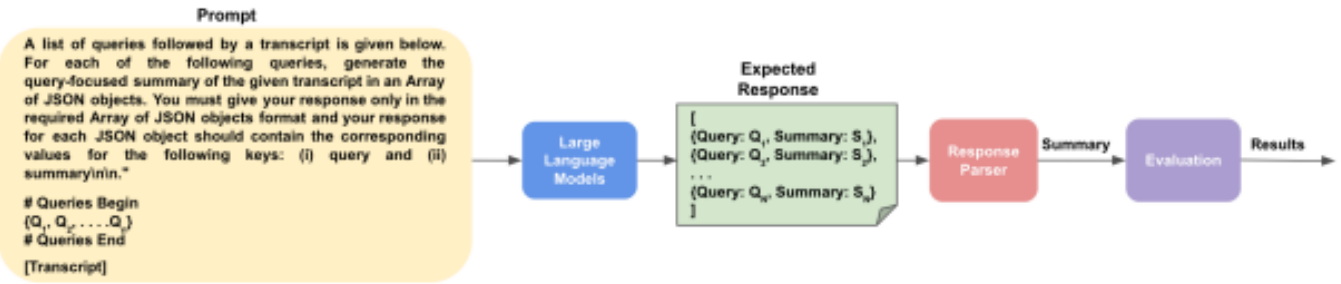

Figure 10: Prompt GenAI System. Source: (Laskar et al., 2024).

## IV. E-GENAI & e-AI

Enterprise Generative Artificial Intelligence (E-GenAI) involves the use of Large Language Models (LLMs) and deep neural networks specialized in generating content such as text, code, images, voice, videos, and other forms of data. It is applied to tasks like creating product descriptions, technical documentation, and reports; summarizing documents; personalizing customer content; producing media for marketing campaigns; and assisting in software development by writing, refining, or debugging code. E-GenAI is built on two essential components: (1) prompt engineering, which converts user inputs into creative and contextually appropriate outputs for business purposes, and (2) LLMs and deep neural networks, which specialize in content generation, producing original outputs based on these prompts to fulfill enterprise requirements. Haki & Safaei (2024) explore the impact and evolution of GenAI on enterprise digital platforms, focusing on three key dimensions: capabilities, architecture, and governance. The rapid advancement of LLMs, such as the Generative Pretrained Transformer (GPT), and other GenAI technologies necessitates a thorough understanding of how these innovations are reshaping the enterprise ecosystem.

Enterprise Artificial Intelligence (e-AI) refers to the application of various AI tools, including data science, machine learning, predictive analytics, robotic process automation, optimization algorithms, and techniques like supervised and unsupervised learning, decision trees, support vector machines, neural networks, and deep learning. These AI tools are used to address a wide range of business problems, such as process optimization, classification, regression, clustering, risk assessment, and customer analytics. e-AI is primarily focused on tasks like automation, data processing, and decision support. It helps identify patterns, analyze data, and provide actionable insights. For instance, e-AI can automate workflows (e.g., CRM or supply chain optimization), classify data (e.g., labeling images), group data (e.g., segmenting customers based on behavior), predict outcomes (e.g., sales forecasting or fraud detection), and support decision-making (e.g., through data analysis, anomaly detection, and process optimization).

These AI systems are trained on large datasets to perform tasks that enhance business outcomes, such as automating processes, making predictions, or supporting decision-making. The emphasis is on analyzing and interpreting historical data to drive process automation and support decision-making process, rather than generating new content (GenAI). The main goal of e-AI is to extract value from the enterprise's existing data and processes.

According to Enterprise-knowledge, a consulting company in the areas of knowledge and information management, software development, and project management, e-AI goes beyond data science, machine learning, and other AI tools. It focuses on solving clearly defined business problems within a specific organizational context, using reliable data, a variety of skill sets, and ensuring that the solution is explainable (e-AI, 2024).

The sociotechnical approach that we propose for the E-GenAI can enhance collaboration within the Providers–Enterprise–Customers ecosystem by employing various LLMs to analyze business information in two phases. First, it utilizes GenAI-based SCM–ERP–CRM platforms. Second, it leverages GenAI-based BI–FL–TRIZ–KM–IKM platforms. Generative AI (GenAI) is reshaping business reality and creating a new ecosystem. This evolution transitions from a vision of the enterprise integrating business information flows (production of products and services, the OID model) to one that incorporates business knowledge flows (production of knowledge capital (assets), the OIDK model). Ultimately, it leads to an Enterprise Generative Artificial Intelligence (E-GenAI) ecosystem based on LLMs.

### A. LLMs

Large Language Models (LLMs), such as ChatGPT, have amazed users with their capability to generate text that is nearly indistinguishable from that written by humans (Zhao et al., 2023). LLMs are deep neural networks trained on extensive language data using self-supervised learning methods. GPT (Generative Pre-trained Transformer) and its successors, including GPT-2, GPT-3.5, GPT-4o and the latest GPT o1, developed by OpenAI, have transformed the field of Natural Language Processing (NLP) with their ability to generate high-quality and coherent text. GPT is a generative model, meaning it can autonomously generate text by predicting the next word in a given sequence.
Thus, LLMs serve as the foundation for new GenAI systems. These GenAI-powered response engines, such as ChatGPT, Gemini, Copilot, LLaMA, Claude 3, Grok, AlphaCode, Copy,

Rephrase, Pi, Perplexity, and others, deliver accurate, reliable, and real-time answers to any query. They contribute to the creation of a new business ecosystem for B2C, B2B, and B2G organizations. For example, ChatGPT and Rephrase help generate personalized content for customer emails, creating more engaging and natural interactions. Copy, Pi, and Perplexity boost creativity and support the development of innovative products and services. AlphaCode and Copilot assist in automating repetitive task workflows.

The rapid advancement of these AI technologies, necessitates a thorough understanding of how these innovations are reshaping the enterprise ecosystem.

Singh et al. (2024) investigate whether GenAI technology could enhance future organizational performance. Their model identifies six key dimensions: the adoption of generative AI, environmental dynamism, exploratory innovation, exploitative innovation, ethical dilemmas, and organizational performance.

Kanbach et al. (2024) explore how E-GenAI and e-AI influence business model innovation (BMI) across industries, particularly in value creation innovation, new proposition innovation, and value capture innovation.

Haki & Safaei (2024) analyze the impact and evolution of GenAI on enterprise digital platforms, focusing on three core areas: capabilities, architecture, and governance.

Solaiman et al. (2024) provide a comprehensive view of AI systems' societal impact, focusing on their measurable and potentially harmful effects. They examine two areas: (1) the impact of GenAI on issues such as bias, misinformation, intellectual property, and content authenticity in content generation, and (2) the effect of e-AI on fairness, transparency, and accuracy in data analytics, emphasizing ethical, unbiased, and reliable decision-making.

Mitra et al. (2024) study the potential application of LLMs for information access applied in information retrieval (IR) systems using a sociotechnical approach.

Figure 11 illustrates the E-GenAI business ecosystem that we propose, which integrates GenAI-based platforms for SCM, ERP, and CRM with GenAI-based platforms for BI, FL, TRIZ, KM, and IKM.

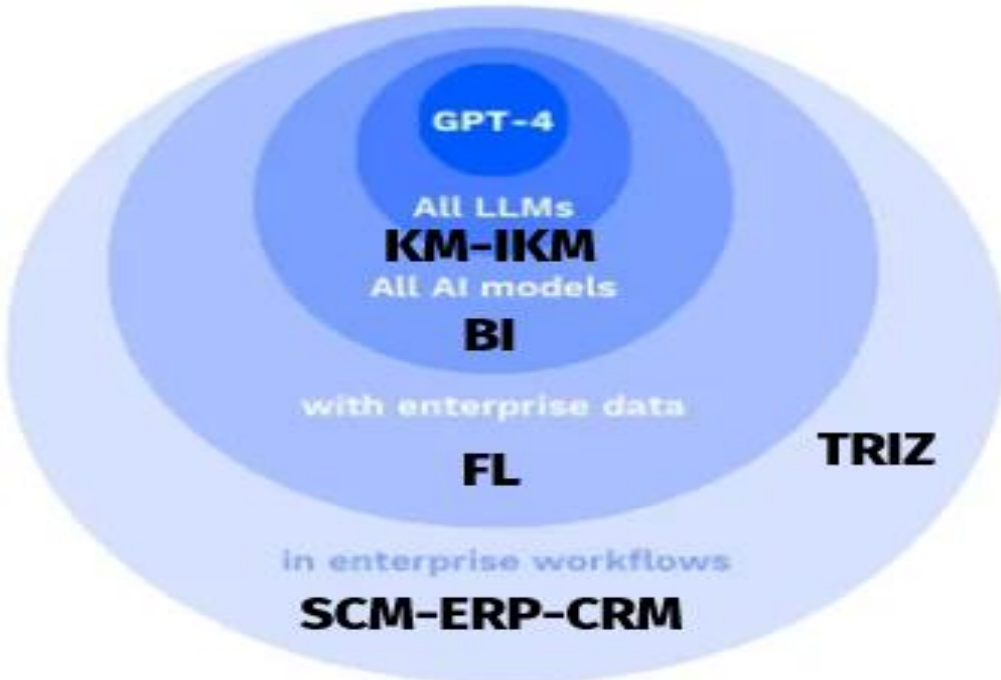

Figure 11: E-GenAI Business Ecosystem. GenAI-based SCM–ERP–CRM platforms and GenAI-based BI–FL–TRIZ–KM–IKM platforms. Source: adapted from (Dilmegani, 2024).

The E-GenAI business ecosystem is built on the AIMultiple model, which demonstrates the potential of integrating GPT-4, LLMs, AI models, enterprise data, and enterprise workflows (Dilmegani, 2024).

The sociotechnical approach that we propose, aligns Large Language Models (LLMs) through the E-GenAI (OID) model (see Figure 12).

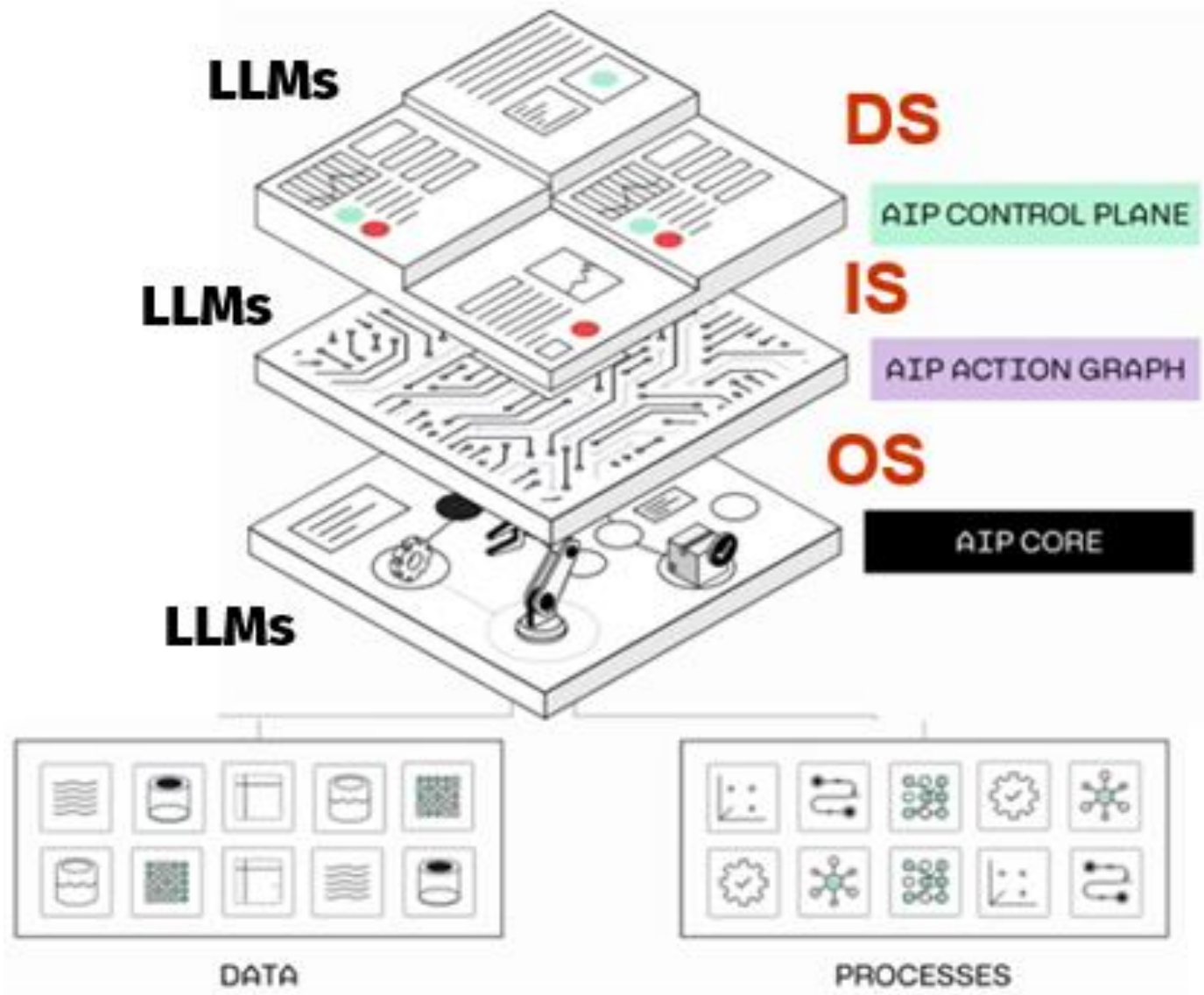

Figure 12: Sociotechnical alignment of LLMs in the E-GenAI (OID) model. Source: adapted from (Dennis, 2024).

The E-GenAI (OID) model is based on the Artificial Intelligence Palantir (AIP) model and consists of three levels. Level 1: AIP Core, which handles data and processes, enabling LLMs to operate on the enterprise private network with a well-optimized data foundation. Level 2: AIP Action Graph, which governs the LLMs, specifying when they can operate independently and when human input is required. Level 3: AIP Control Plane, which controls and manages LLM activity (AIP, 2024). Palantir's model emphasizes key concepts, making the underlying levels more explicit, including how they are grouped, applied LLMs, and managed by different teams. This allows Palantir engineers to better align products around shared concepts, adapt those concepts to meet user needs, and improve communication with non-technical teams within the company (Wilczynski et al., 2023).

In the E-GenAI (OID) model, humans handle the labeling or tagging of objects, and then the machine (such as ChatGPT) learns from these human inputs to determine the correct labels or tags (F8, 2024). This process resembles a finite automata (see Figure 13), where humans perform tasks uniquely suited to their capabilities (labeling or tagging objects), while generative artificial intelligence (GenAI) excels in tasks suited to algorithms (learning to label or tag objects automatically). Additionally, ChatGPT and other LLMs have the ability to self-learn and adapt using all available GPT models.

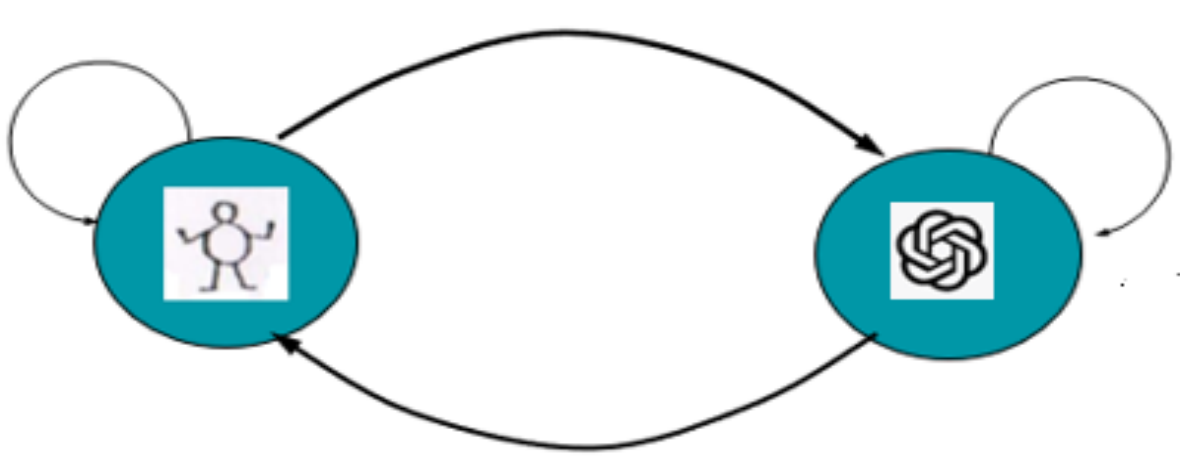

Figure 13: Finite Automata: Human–Machine. Source: main author.

### B. LLMs and Social Media

Applying the sociotechnical approach to a social media platform, the social component consists of users (both individuals and enterprises). They interact with each other through likes, photos, videos, and comments. Figure 14 illustrates a graph depicting these user interactions.

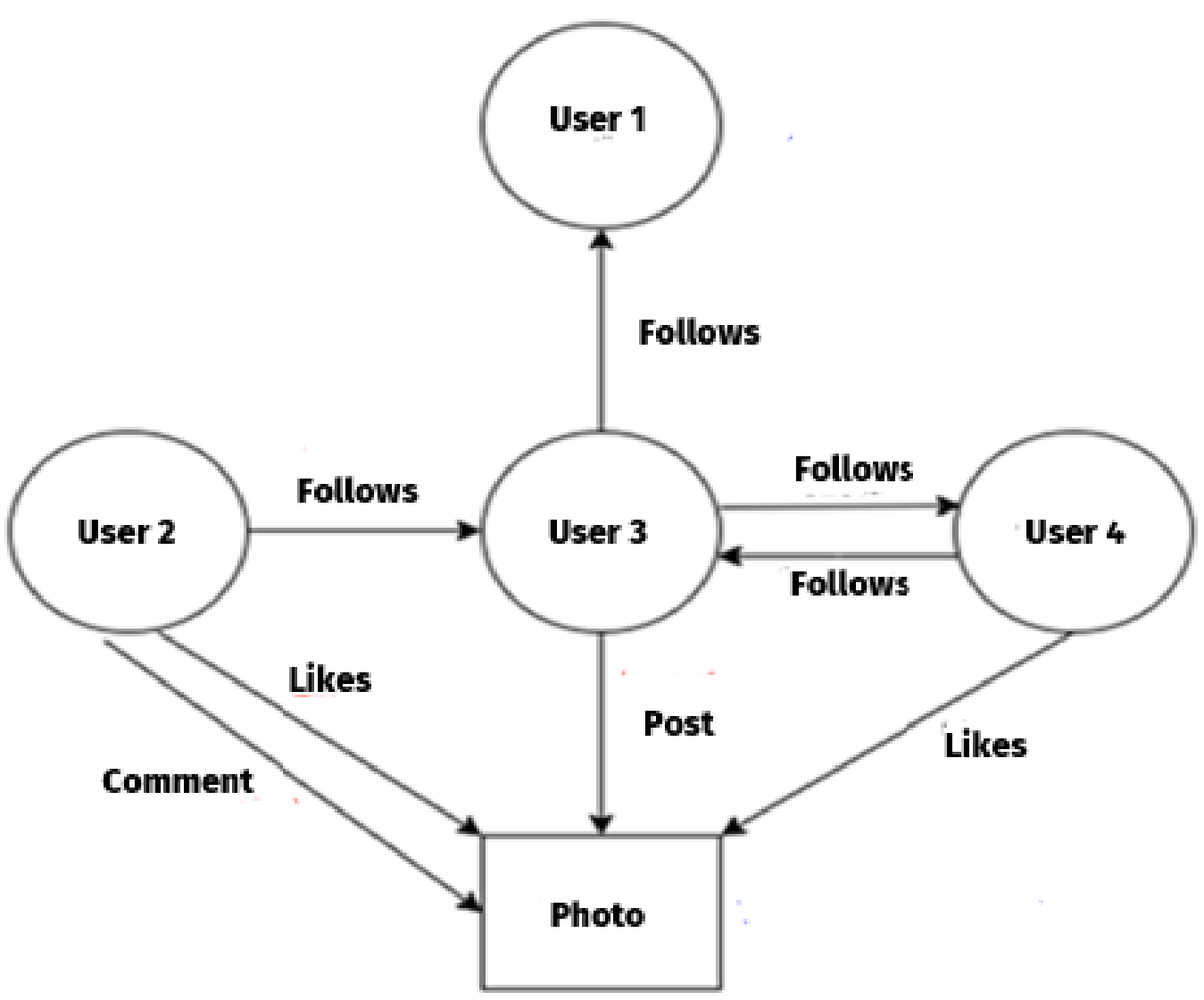

Figure 14: Users Ineraction Graph. Source: main author.

User 3 posts a photo, User 4 likes it and adds a comment, while User 4 only leaves a like. Users can choose to view a post, like it, comment on it, or ignore it entirely (as in the case of User 1). Their actions are driven by the content of the post and their interest. This interaction model represents a complex system. Now, the technical component of the sociotechnical approach can be illustrated using an Entity Relationship Diagram (see Figure 15).

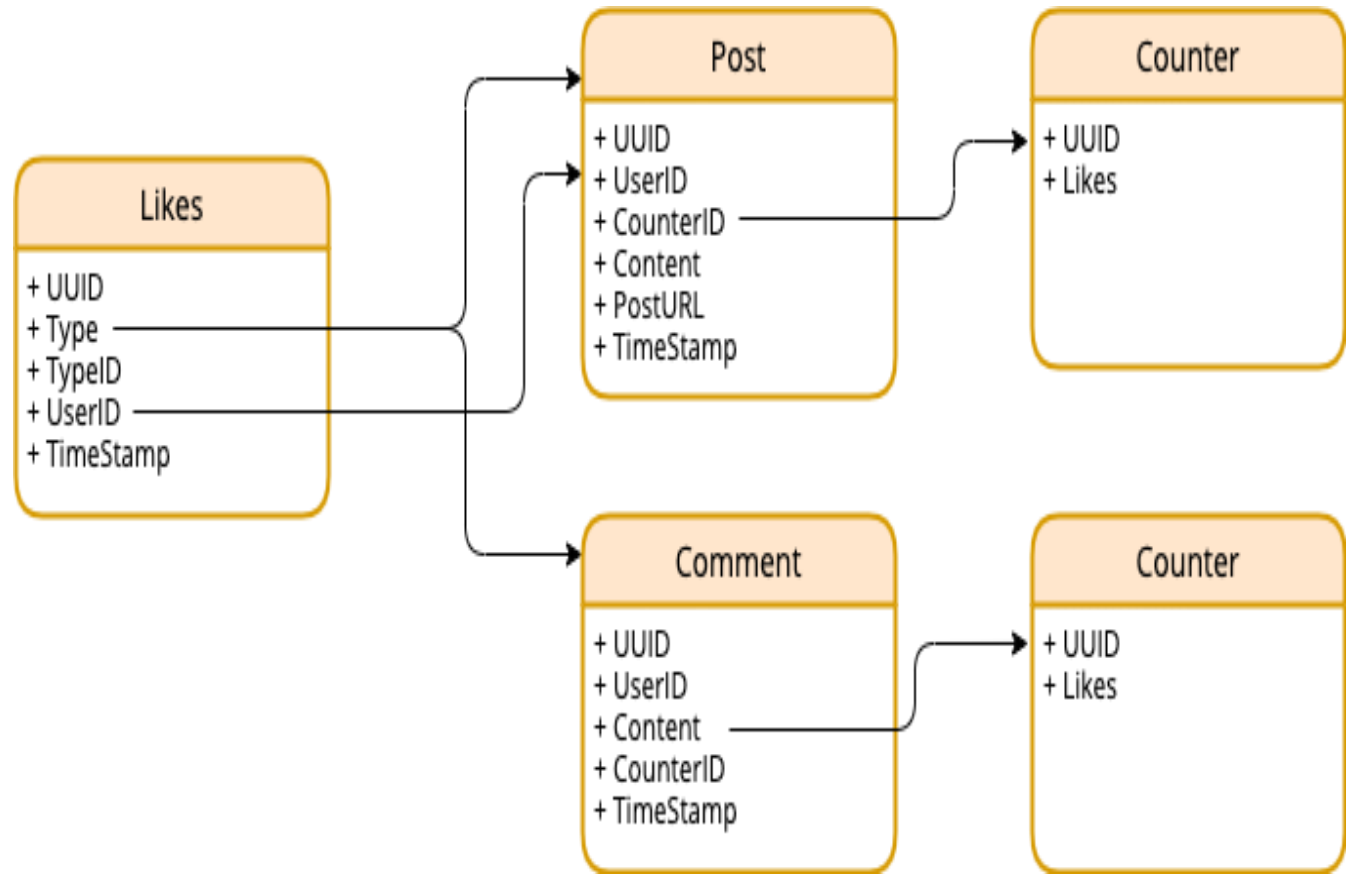

Figure 15: Entity Relationship Diagram. Source: (Hudakova, 2024).

Meanwhile, Figure 16 illustrates a relation called "Binding" between a User and a Followee. A User is someone who follows others, while a Followee is someone being followed. To find all followers of a specific user, the query would be SELECT User FROM Binding WHERE Followee = userID. Conversely, to find all users that a particular user follows, the query would be SELECT Followee FROM Binding WHERE User = userID.

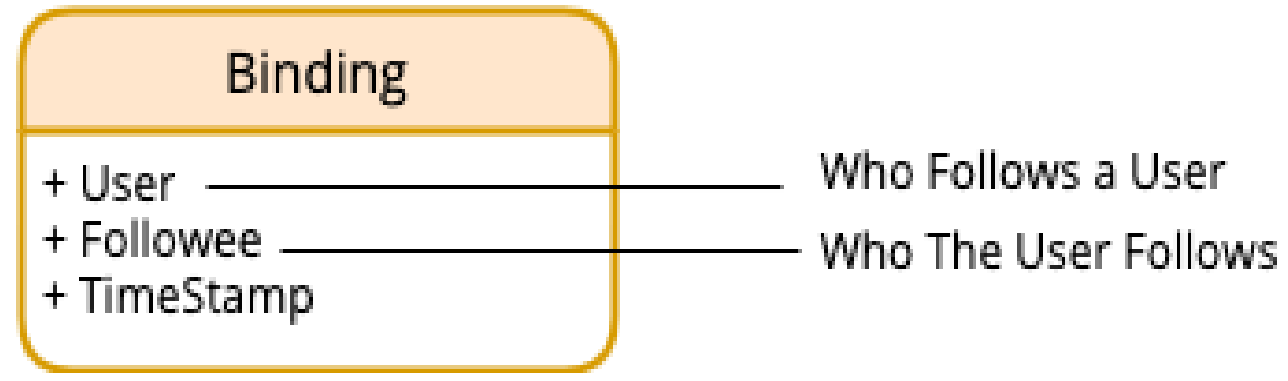

Figure 16: Binding Entity. Source: (Hudakova, 2024).

To understand the dynamic of user interactions, Figure 17 (left) illustrates the relationships between Followers, Posts, and Followees. A Follower is a user who follows another, while a Followee is someone whose posts are followed by others. Figure 17 (right) presents a finite automata with two states: "Follows" and "Followees," along with four transitions: {Follows/Follows, Follows/Followees, Followees/Followees, Followees/Follows}. This finite automata depicts the interactions between Followers and Followees, and recursively, the interactions between each user and their respective Followers and Followees.

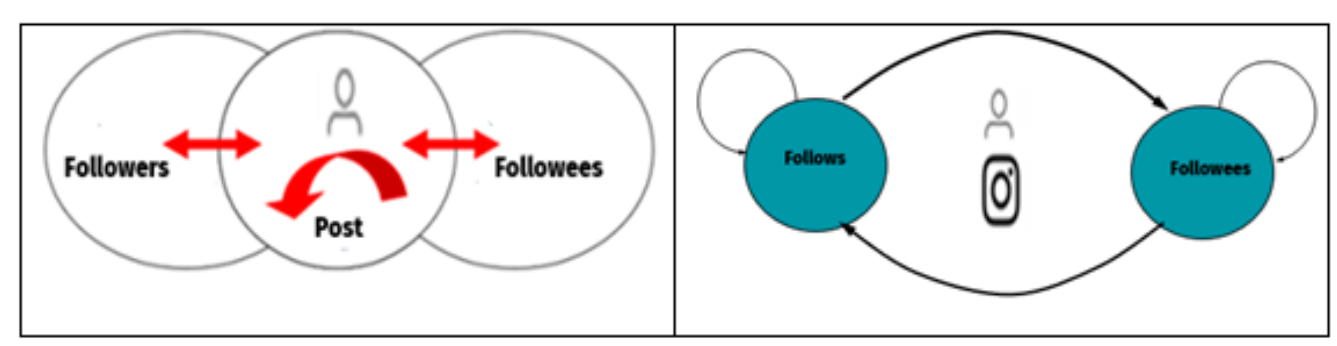

Figure 17: Left: Followers–Post–Followees relationships. Right: Finite Automata: Follows–Followees. Source: main author.

This finite automata enables the construction of LLMs to understand specific characteristics of users' Followers and Followees. For example, do all Followers share the same traits? Identifying, storing, and classifying these characteristics presents a business opportunity. LLMs can analyze this information to design advertising campaigns tailored to users' preferences. Creating new content is crucial for attracting new Followers, with success depending on the content's relevance and the users' interests. A finite automata processes a sequence of inputs (in this case, user actions) and moves through different states depending on those inputs. This can help identify and track patterns in Followers–Followees relationships, such as how users engage with content, how Followers interact with specific Followees, or how communities form around certain users. The E-GenAI (OID) model uses finite automata to understand the relationships between Followers and Followees in social media, allowing LLMs to analyze user behavior. This analysis could be leveraged for business opportunities, like designing targeted marketing campaigns based on user preferences. Finite automata can be a powerful tool for modeling user interactions on social media platforms, particularly for understanding relationships between Followers and Followees.

### C. LLMs and TRIZ

The Theory of Inventive Problem Solving (TRIZ) is a powerful methodology widely used in the engineering community to address complex problems by identifying and resolving contradictions (Cavallucci, 2017). It involves recognizing a contradiction—where improving one feature negatively impacts another—and using the 39 TRIZ engineering parameters to select one or more of 40 inventive principles from the TRIZ contradiction matrix to develop an innovative solution to the original problem. Figure 18 outlines the four steps of problem-solving using TRIZ: Step 1: Define the Problem; Step 2: Identify the Contradiction; Step 3: Apply TRIZ Inventive Principles; and Step 4: Implement the Solution.

Let's see an example of how TRIZ can be applied in the OID Model. A retail company faces a problem between improving operational efficiency (OS) by maintaining high inventory levels to avoid stockouts and reducing inventory costs (DS) by keeping stock levels low. This creates a contradiction: high inventory increases operational efficiency but raises costs, while low inventory reduces costs but risks stockouts and lost sales. Steps to apply TRIZ in the OID Model:

Step 1: Define the Problem (Operational System - OS). The retail company aims to improve operational efficiency by maintaining high stock levels to meet customer demand. However, increasing inventory raises costs, tying up capital and increasing storage costs.

Step 2: Identify the Contradiction (Decision System - DS). Contradiction: The desire to have high inventory levels for efficiency conflicts with the need to reduce costs by minimizing inventory.

Step 3: Apply TRIZ Principles. Using the TRIZ contradiction matrix (improving feature vs. worsening feature), this problem can be categorized as a conflict between improving system efficiency (TRIZ parameter #39–Productivity) and reducing cost (TRIZ parameter #28–Measurement accuracy). The TRIZ matrix [1] suggests four TRIZ inventive principles to address this conflict: TRIZ principle #10–Prior Action: Pre-position inventory in strategic locations (closer to high-demand areas) before peak periods to reduce restocking delays without overstocking everywhere. TRIZ principle #1–Segmentation: Split the inventory into smaller, more manageable parts. For example, apply a Just-in-Time (JIT) approach for slower-moving products while keeping higher stock levels for fast-selling items. Finally, TRIZ principle #34–Discarding and recovering and TRIZ principle #28–Mechanics substitution do not apply to the problem of operational efficiency vs. inventory costs in a retail company.

Step 4: Implement the Solution. The operational system (OS) now pre-positions inventory in key locations (urban areas with high demand) before peak sales periods while using a leaner JIT inventory strategy in lower-demand regions. The decision system (DS) evaluates these strategies and confirms that the JIT system helps reduce costs while maintaining service levels in high-demand.

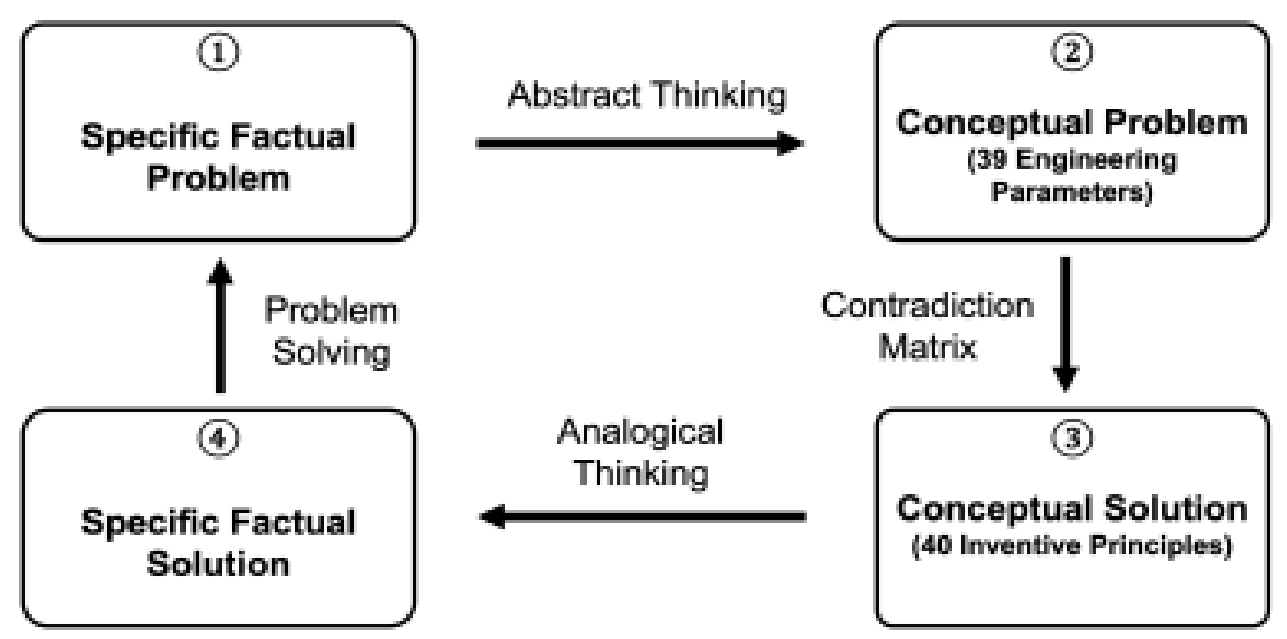

Figure 18: TRIZ Problem-Solving Process. Source: (Jiang & Luo, 2024).

Seraia and Seryi (2016) present the 39 TRIZ engineering parameters and the 40 TRIZ inventive principles, illustrating them with scientific and industrial applications.

Jiang & Luo (2024) explore the latest advancements in large language models (LLMs) for a generative approach to TRIZ. They introduce AutoTRIZ, an AI-driven ideation tool that utilizes LLMs to automate and improve the TRIZ problem-solving methodology (see Figure 18). AutoTRIZ takes a problem statement defined by the user and generates TRIZ-based solutions by leveraging the TRIZ contradiction matrix. Figure 20 presents the framework of AutoTRIZ as proposed in their study.

---

[1] https://www.triz40.com/aff_Matrix_TRIZ.php
https://www.triz40.com/TRIZ_GB.php

Trapp & Warschat (2024) propose a method for extracting TRIZ contradictions from text using GPT-4. Their approach includes outlining the problem statement, detecting contradiction parameters (improving feature vs. worsening feature), identifying relevant inventive principles, and generating solutions through prompt engineering. Figure 19 illustrates a similar problem statement prompt from user with AutoTRIZ, which involves defining the problem statement, detecting engineering contradictions, identifying inventive principles, and generating solutions.

**Problem Statement**

We need to resolve the wear and damage issues caused by the transition to metal shots in a system of plastic piping originally intended for plastic pellets.

**Detected Engineering Contradiction**

**Improving Feature: Area of stationary object (Index 6):** The surface area of the plastic piping, which needs to withstand the impact and wear caused by the metal shots.
**Worsening Feature: Stability of the object's composition (Index 13):** The integrity and wholeness of the plastic piping, affected by wear and damage from the metal shots.

**Identified Inventive Principles**

**Extraction (Index 2):** Extraction involves removing or separating an undesirable or disturbing part from an object or system. It can also include extracting the desirable part or isolating the critical part of a system. This principle can be applied to separate the damaging effect of the metal shots from the plastic piping.
**Strong oxidants (Index 38):** This principle involves enhancing the reaction with oxygen by using materials that combine with oxygen readily or adding more oxygen to the system. It can be applied to enhance the resistance of the plastic piping to the impact and wear caused by the metal shots.

**Generated Solutions**

**Solution 1 - Redesign of Elbows Using Extraction Principle**
- Applied Principle: Extraction (Index 2)
- Detailed Solution Description: Redesign the elbows of the plastic piping to include a separator mechanism or structure that can isolate the metal shots from directly impacting the plastic material. This separation mechanism can redirect the metal shots away from the vulnerable areas of the piping, thereby minimizing wear and damage.

**Solution 2 - Surface Treatment with Strong Oxidants**
- Applied Principle: Strong oxidants (Index 38)
- Detailed Solution Description: Apply surface treatment to the plastic piping using materials that combine readily with oxygen, enhancing the piping's resistance to wear and impact. This could involve incorporating a protective layer or coating on the piping's surface to effectively mitigate the damaging effects of the metal shots.

Figure 19: Example of problem statement prompt from user using AutoTRIZ. Source: (Jiang & Luo, 2024).

The engineering problem statement of Figure 19 is about the pneumatic transportation of metal shots through a system of plastic piping. The user original statement is: "We are faced with a challenge involving the pneumatic transportation of metal shots through a system of plastic piping originally intended for plastic pellets. The transition to metal shots, despite their advantages for production purposes, has led to significant wear and damage, particularly at the pipe's elbows. This issue arises from the incompatibility between the metal shots and the existing plastic elbow design. The task is to identify and implement a solution that resolves this conflict, ensuring the system's durability and effectiveness for transporting metal shots" (Jiang & Luo, 2024).

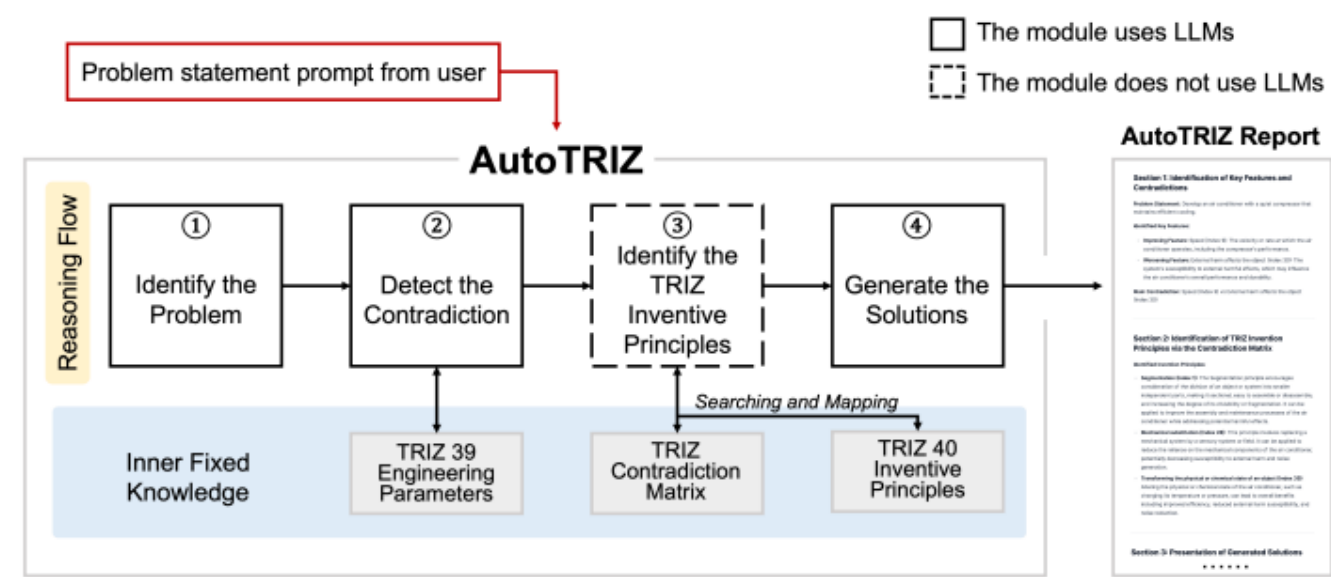

Figure 20: Problem-Solving Process AutoTRIZ. Source: (Jiang & Luo, 2024).

Chen et al. (2024) examine the application of LLMs, specifically GPT-4, within the TRIZ framework. They demonstrate that solutions generated through this approach enhance traditional problem-solving processes (see Figure 18) by tapping into extensive knowledge bases and advanced reasoning capabilities to produce innovative solutions. Figure 21 outlines the TRIZ-GPT framework proposed in their research.

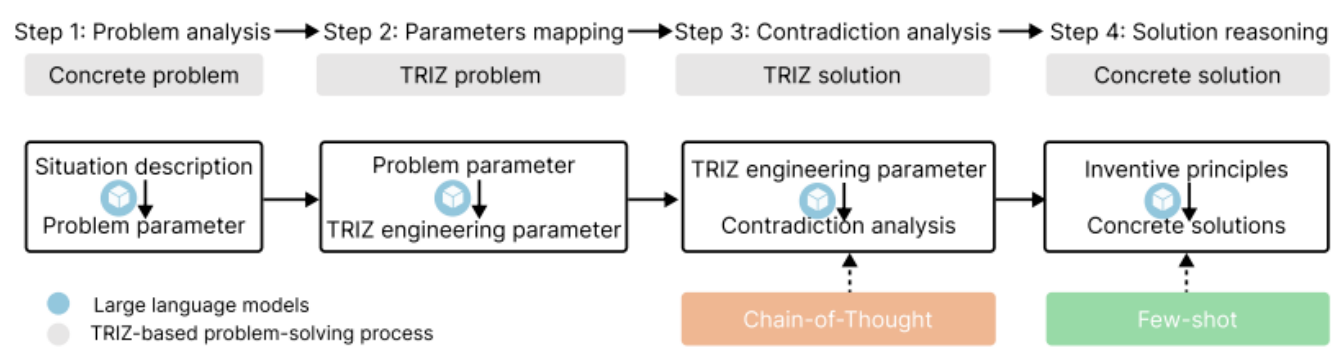

Figure 21: Problem-Solving Process TRIZ-GPT. Source: (Chen et al., 2024).

## CONCLUSIONS

The paper proposes a sociotechnical approach to analyze a business ecosystem of data, information, and knowledge, using Le Moigne's OID and Ermine's OIDK model. Le Moigne's OID model enables us to describe the enterprise as a living system that evolves, organizes, and structures itself through its relationships with Providers, Enterprise, and Customers via platforms like SCM, ERP, and CRM. We align with the OID model by using Fuzzy Logic (FL) to handle imprecise, uncertain, or incomplete data within the information system, Business Intelligence (BI) to achieve the strategic objectives of the decision system, and TRIZ to address complex problems within the operational system. Ermine's OIDK model incorporates a knowledge component into the OID model. We align with the OIDK model by using Knowledge Management (KM) to capitalize on knowledge assets, and Imperfect Knowledge Management (IKM) to handle the imprecise, uncertain, or incomplete aspects of competence or cognition flows.
The primary goal of this theoretical paper is explore an E-GenAI business ecosystem. This ecosystem integrates GenAI-based platforms for SCM, ERP, and CRM with those for BI, FL, TRIZ, KM, and IKM, in order to align Large Language

Models (LLMs) within the framework of the E-GenAI (OID) model. The social media platform functions as an integrated system designed to create value through social interactions between people and enterprises. To facilitate the development of LLMs capable of identifying specific user characteristics, we have defined these interactions using finite automata.

In future research, we plan to utilize the E-GenAI (OID) model to examine the dynamics and context-sensitive capabilities of LLMs within a specific business context. This will allow us to consider including case studies or empirical data that demonstrate the effectiveness of our proposed model in Real-Time Enterprise settings. We address more explicitly the limitations and challenges associated with implementing the E-GenAI (OID) model, especially concerning technological infrastructure, data privacy, and ethical considerations. We'll create a Finite Automata States and Transitions to modeling (1) Human – Machine (humans handle labeling and GenAI like ChatGPT refine these labels) for automating customer feedback classification, and (2) to modeling Followers – Followees for automating user preferences on social media platforms.